\documentclass[twocolumn,amsmath,amssymb,prl]{revtex4}
\usepackage{graphics,graphicx}
\usepackage{braket}
\begin{document}
\title{Local spin ice order induced planar Hall effect in Nd-Sn artificial honeycomb lattice}
\author{J. Guo$^{1}$}
\author{G. Yumnam$^{1}$}
\author{A. Dahal$^{1}$}
\author{Y. Chen$^{1}$}
\author{V. Lauter$^{2}$}
\author{D. K. Singh$^{1,*}$}
\affiliation{$^{1}$Department of Physics and Astronomy, University of Missouri, Columbia, MO 65211, USA}
\affiliation{$^{2}$Neutron Science Directorate, Oak Ridge National Laboratory, Oak Ridge, TN 37830, USA}
\affiliation{$^{*}$Corresponding Author: singhdk@missouri.edu}

\begin{abstract}
Geometrically frustrated materials, such as spin ice or kagome lattice, are known to exhibit exotic Hall effect phenomena due to spin chirality. We explore Hall effect mechanism in an artificial honeycomb spin ice of Nd--Sn element using Hall probe and polarized neutron reflectivity measurements. In an interesting observation, a strong enhancement in Hall signal at relatively higher temperature of $T$ $\sim$ 20 K is detected. The effect is attributed to the planar Hall effect due to magnetic moment configuration in spin ice state in low field application. In the antiferromagnetic state of neodymium at low temperature, applied field induced coupling between atomic Nd moments and conduction electrons in underlying lattice causes distinct increment in Hall resistivity at very modest field of $H$ $\sim$ 0.015 T. The experimental findings suggest the development of a new research vista to study the planar and the field induced Hall effects in artificial spin ice.
\end{abstract}

\maketitle

Hall effect probe is one of the quintessential methods to derive information about electric charge carriers and their polarity in solid state materials.\cite{Lindberg,Karplus} In magnetic materials with chiral spin arrangement, as found in geometrically frustrated bulk spin ice compound or skyrmion lattice,\cite{Harris,Pappas,Taguchi,Neubauer,Ong} an additional contribution to Hall  resistivity arises.\cite{Taguchi} The additional term is solely dependent on the intrinsic net spin chirality due to the sub-lattice magnetization, given by $S_i\cdot (S_j\times S_k)$ where $S_i$, $S_j$ and $S_k$ are local spins forming a chiral loop.\cite{Nagaosa} Electric charge carriers gain a net phase, called Berry phase, while traversing the chiral loop, reflecting as an extra contribution in the typical Hall resistance measurement.\cite{Ye,Nagaosa2} More recently, thermal anomalous Hall effect was also proposed in two-dimensional quantum material.\cite{Lee} Aside from the anomalous and thermal Hall effects, underlying magnetism due to the spin chiral anomaly or magnetic order can induce the novel planar Hall effect where the Hall voltage is coplanar to net magnetization.\cite{Potter,Burkov} Typically, the planar Hall effect is prominent in two-dimensional materials with ferromagnetic and topological characteristics or, in bulk materials where these properties are primarily confined to the surface.\cite{Nitesh,Nandy,Ogrin} The planar Hall effect (PHE) is given by the following anisotropic magnetoresistance:\cite{Potter,Ogrin}
\begin{equation}
E = (\rho_{\parallel}-\rho_{\perp})\boldsymbol{m}(\boldsymbol{j\cdot m})+\rho_{H}(\boldsymbol{m}\times\mathbf{j}) \label{eq:total_E}
\end{equation}
where $\boldsymbol{j}$ and $\boldsymbol{m}$ are current density and unit magnetization vector given by \textbf{M}/$\mid$\textbf{M}$\mid$, $\rho_{\perp}$ and $\rho_{\parallel}$ are perpendicular and parallel resistivities with respect to $\boldsymbol{j}$ and $\rho_H$ is the normal Hall resistivity. The first and second terms in the above expression denote the planar and normal Hall effects, respectively.\cite{Potter,Ogrin} Thus, a non-zero contribution to the Hall resistivity can be obtained solely due to the coupling between current density and magnetization vector. Conversely, the planar Hall effect can be used to detect magnetic anisotropy in two dimensional materials.\cite{Tang} 

Artificial spin ice provides a niche platform to investigate the phenomenon of planar Hall effect at relatively higher temperature.\cite{Chern} Additionally, the disorder-free environment in nanostructured geometry paves way for the elucidation of intrinsic properties in the system. In this report, we discuss Hall effect measurements on neodymium based artificial honeycomb spin ice. We observe strong signature of planar Hall effect due to local spin ice order at moderately high temperature of $T \sim$ 20 K. The honeycomb lattice is made of 3 nm thick neodymium connecting elements with typical length and width of 12 nm and 5 nm, respectively. Neodymium elements are preceded by a metallic buffer layer of 3 nm thick tin, in clean contact with Nd layer, in honeycomb construction, see Fig. 1a. Although bulk Nd is antiferromagnetic at low temperature $T < T_N$ $\simeq$ 18 K, it tends to be ferromagnetic for $T > T_N$ in thin film or amorphous composition.\cite{Bak,Thomale,Taylor,Dai} The honeycomb lattice, made of thin film of Nd, is expected to follow similar magnetization behavior. Above $T_N$, magnetic moment aligns along the length of connecting element due to the shape anisotropy.\cite{Heyderman,Shen,Mengotti} Consequently, two types of local moment arrangements emerge on honeycomb vertices: `two-in \& one-out' (or vice-versa) configuration, characterized by two incoming and one outgoing magnetizations, or 'all-in or all-out' state where all three moments are either pointing in or pointing out on a given vertex. At low temperature, the lattice is mostly populated by `two-in \& one-out' (or vice-versa) moment arrangement. The energetic `all-in or all-out' type defects may also be present, but the density must be very small due to high energy cost.\cite{Schiffer} The `two-in \& one-out' state, identified as the ice-rule, naturally manifests a local chirality in moments arrangement.\cite{Branford} The large moment size of individual element (of the order of 10$^{3}$ $\mu_B$) makes long range dipolar magnetic interaction as the dominant term in the Hamiltonian.\cite{Schiffer} Therefore, the local chirality due to the short-range spin ice order is preserved across the honeycomb lattice at modest temperature. When current is applied to honeycomb lattice, then the current density $\boldsymbol{j}$ is either parallel or antiparallel to magnetization $\boldsymbol{M}$ along honeycomb element, see schematic Fig. 1c. Since, \textbf{\textit{j}} is strictly along the length of the honeycomb element, Eq. (1) yields a finite PHE of $E = \Delta \rho. j$ where $\Delta \rho$ = $\rho_\parallel$ - $\rho_\perp$. Modest perpendicular magnetic field application is not sufficient to exert enough torque to alter local moment structure. However, magnetic field application or thermal fluctuation, larger than the shape anisotropy energy, can unsettle the moment along the length of connecting element, thus forbids the development of PHE. Similarly, at $T <$ $T_N$, Nd spins of opposite polarity in the AFM state cancel net moment formation along the honeycomb element. Hence, spins can align to magnetic field application direction. Correspondingly, electric charge carriers in Nd-Sn layers are subjected to a net magnetic field perpendicular to the honeycomb film. This scenario renders the possibility of normal Hall effect, given by $\rho_H.j$, as envisaged in the second term of Eq.(1). The process is schematically described in Fig.1d. 

\begin{figure}
\centering
\includegraphics[width=8.9 cm]{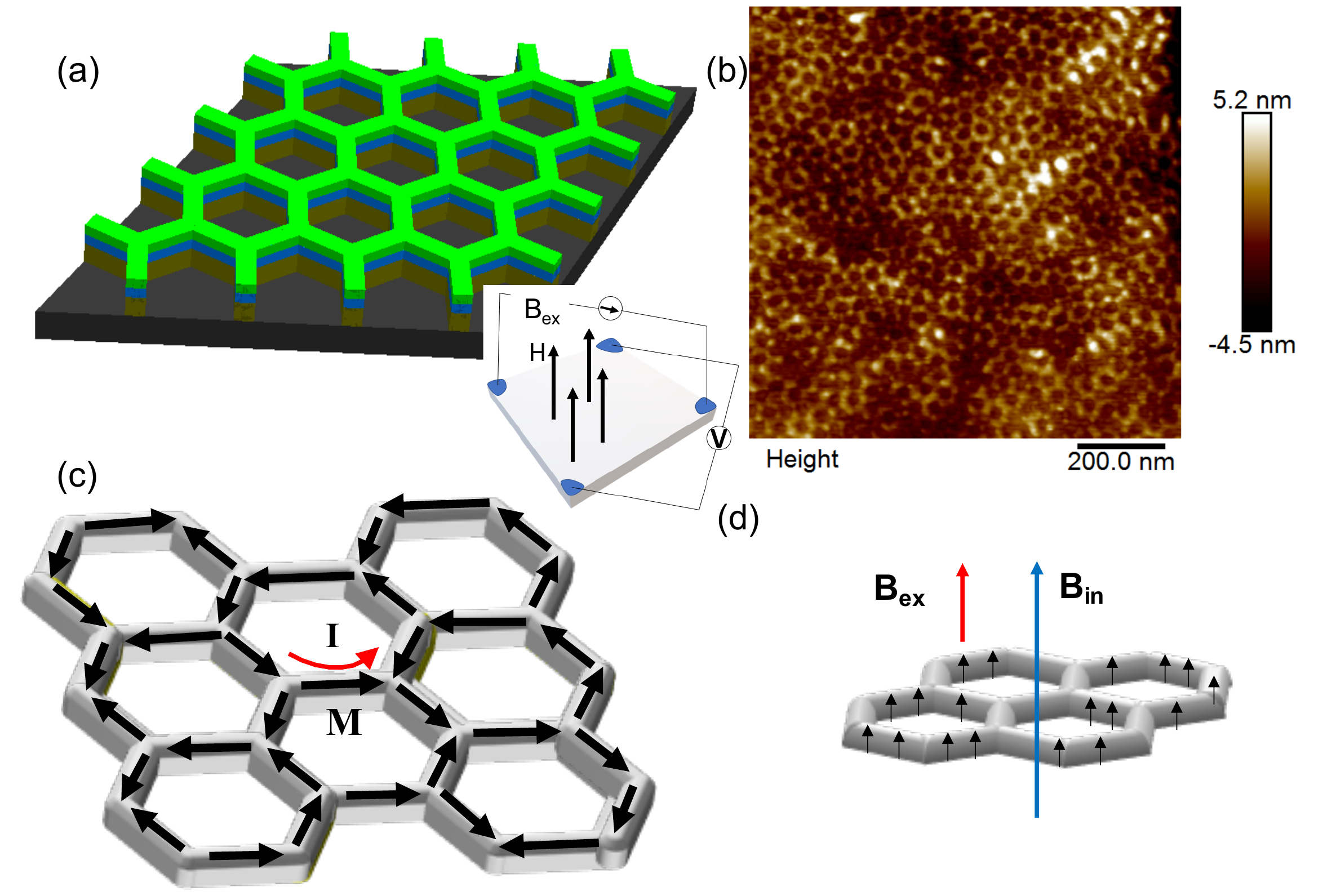} \vspace{-6mm}
\caption{(color online) Nd-Sn honeycomb lattice and illustration of moment configuration due to spin ice state. (a) Schematic design of Nd-Sn artificial honeycomb lattice construction. Here, Nd layer (green) is grown on top of Sn (blue) buffer layer. (b) Atomic force micrograph of Nd-Sn artificial honeycomb lattice. (c) Schematic illustration of local moment chirality due to short-range spin ice configuration. Magnetic moment aligns along the length of connecting element due to shape anisotropy in honeycomb lattice. Magnetic moments interact via magnetic dipolar interaction. Current direction, parallel to magnetic moment direction, gives rise to the planar Hall effect. (d) Below $T_N$, Nd spins are free to align to applied field direction. Thus, electric charge carriers in Nd-Sn film are subjected to an additional perpendicular field, causing normal Hall effect. Inset shows Hall measurement scheme.
} \vspace{-6mm}
\end{figure}

We create neodymium-based honeycomb lattice samples using diblock template method,\cite{Russell} which results in large throughput sample with ultra-small connecting elements, $\sim$12 nm in length, see Fig. 1b. Details about the nanofabrication procedure can be found somewhere else.\cite{Yiyao} Hall probe measurements are performed on a 2$\times$2 mm$^2$ sample using the standard contact configuration, as prescribed by National Institute of Standards and Technology,\cite{NIST} see inset in Fig. 1. Electrical measurements were performed in a cryogen-free 9 T magnet with a base temperature of $\sim$5 K, using a high-quality resistivity bridge from Linear Research. A 1 sq. inch size sample is used for the polarized neutron reflectometry (PNR) measurements on Magnetism Reflectometer, beam line BL-4A of the Spallation Neutron Source (SNS), at Oak Ridge National Laboratory. PNR measurements utilize the time of flight technique in a horizontal scattering geometry with a bandwidth of 5.6 \AA(wavelength varying between 2.6--8.2 \AA). The beam was collimated using a set of slits before the sample and measured with a 2D position sensitive $^{3}$He detector with 1.5 mm resolution at 2.5 m from the sample. The sample was mounted on the copper cold finger of a close cycle refrigerator with a base temperature of $T=5$ K. Beam polarization and polarization analysis was performed using reflective super-mirror devices, achieving better than 98\% polarization efficiency over the full wavelength band. For reflectivity and off-specular scattering the full vertical divergence was used for maximum intensity and a 5\% $\Delta$$\theta$/$\theta$ $\simeq$ $\Delta$$q_z$/$q_z$ relative resolution in horizontal direction. 

\begin{figure}
\centering
\includegraphics[width=9. cm]{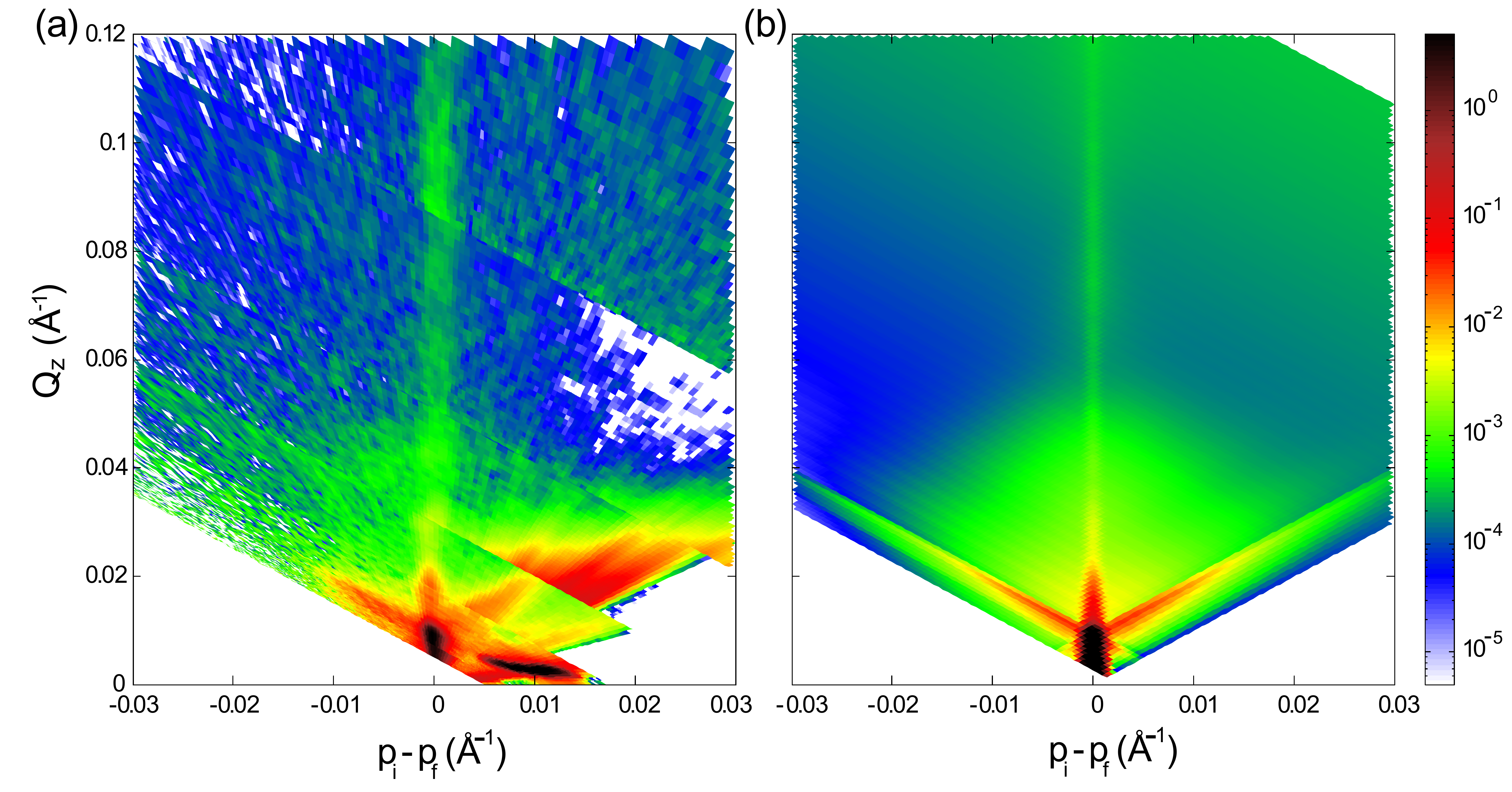} \vspace{-6mm}
\caption{(color online) Elucidating magnetic moment arrangement via polarized neutron reflectometry measurement on Nd-Sn honeycomb. (a) PNR data at $T$ = 25 K. The vertical line at x = 0 represents specular scattering. As we can see, spectral weight is mostly confined to specular reflection. (b) Simulated reflectometry profile, using DWBA modeling, for magnetic moment configuration shown in Fig. 1c. Experimental data is well explained by the simulated profile.
} \vspace{-6mm}
\end{figure}

Evidence to chiral loop arrangement of magnetic moments in neodymium honeycomb lattice is obtained via the analysis of PNR measurements. PNR measurements were performed in a small guide field of $H$ = 0.002 T to maintain the polarization of incident and scattered neutrons. In Fig. 2a, we plot the off-specular intensity measured using spin-up (+) and spin-down (-) neutron at $T$ = 25 K. Here y-axis represents the out-of-plane scattering vector ($q_z= \frac{2\pi}{\lambda}(\sin \alpha_i+ \sin \alpha_f)$) and the difference between the z-components of the incident and the outgoing wave vectors ($p_i-p_f=\frac{2\pi}{\lambda}(\sin \alpha_i- \sin \alpha_f)$) is drawn along the x-axis. Thus, vertical and horizontal directions correspond to the out-of-plane and in-plane correlations, respectively (for detail information, see Lauter et al.). \cite{Lauter} The specular reflectivity lies along the x = 0 line. Experimental data is modeled using the distorted wave Born approximation (DWBA) formulation to understand the nature of magnetic moment correlation.\cite{DWBA,Artur} As shown in Fig. 2b, the numerically simulated reflectometry pattern for magnetic moment configuration, comprised of ‘two-in \& one-out’ or vice-versa states (as shown in Fig. 1c), is found to be in good agreement with the experimental data. The PNR measurements and the associated DWBA modeling illustrates the nature of local moment chirality due to spin ice configuration in Nd-honeycomb lattice, which is crucial to the depiction of planar Hall effect. The nearly equal populations of 'two-in \& one-out' and vice-versa spin ice configurations on honeycomb vertices significantly weaken the net macroscopic magnetization in Nd lattice. However, the DWBA simulation suggests that the two-out \& one-in moment arrangement seems to prevail, albeit weakly, as shown in Fig. 1c.

\begin{figure}
\centering
\includegraphics[width=9.5 cm]{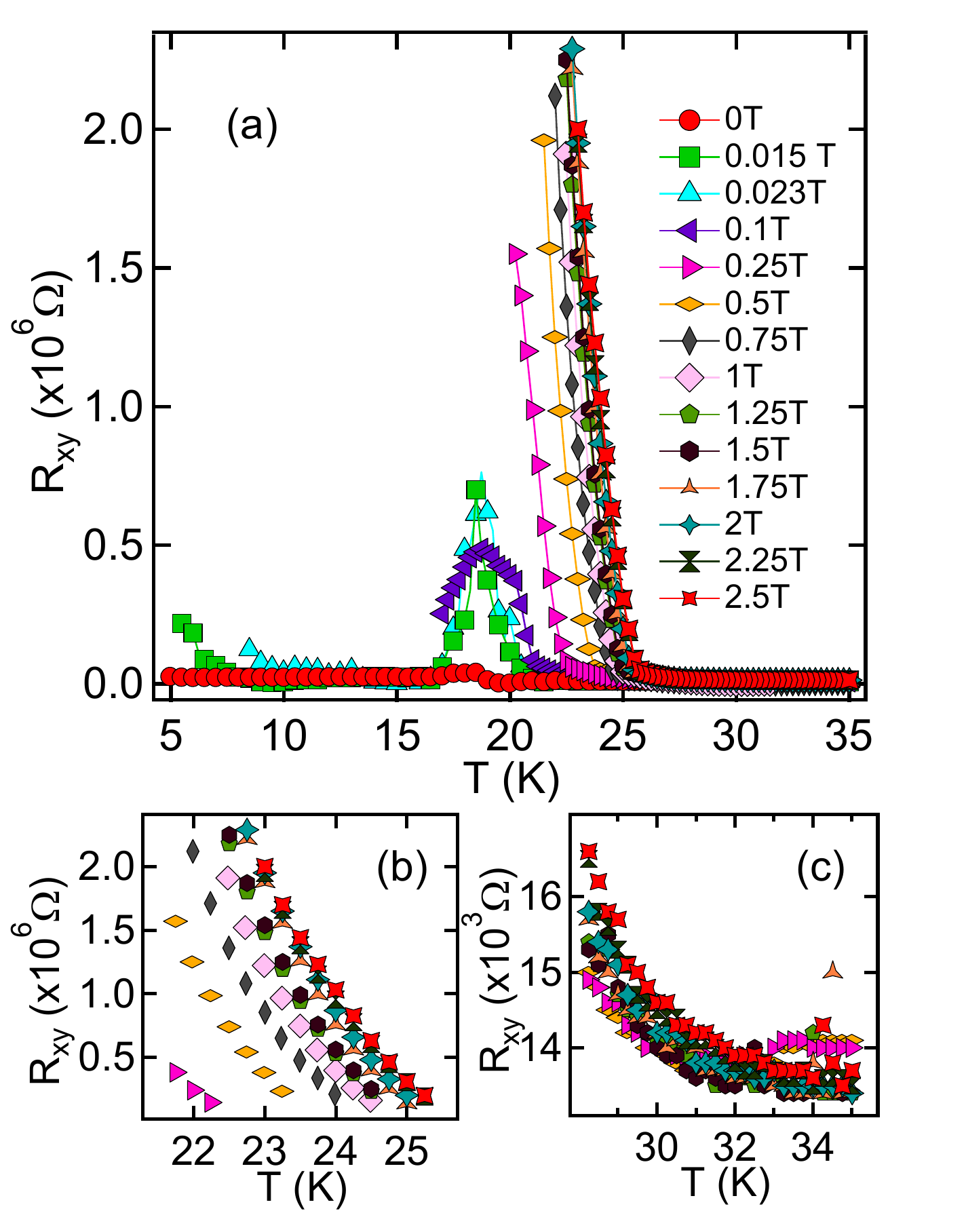} \vspace{-6mm}
\caption{(color online) Plot of Hall resistance ($R_{xy}$) versus temperature at characteristic magnetic fields. (a) Figure shows the plot of $R_{xy}$ vs. $T$(K). (b) $R_{xy}$ dependence on temperature saturates at $H$ = 1.75 T. Magnetic field application of higher magnitude does not affect Hall effect. Rather, experimental data at $H >$1.75 T overlaps on top of each other. (c) Hall effect tends to disappear above $T \sim$ 32 K.
} \vspace{-6mm}
\end{figure}

We show the Hall resistance, $R_{xy}$, as a function of temperature at few characteristic magnetic fields in Fig. 3a. There are several noticeable features in experimental data. First, the Hall resistance manifests strong enhancement around $T\sim$ 20 K. At very low field, $H$ = 0.015 T, $R_{xy}$ increases by more than 100\% with respect to the zero field background as well as the normal resistance $R_{xx}$ when temperature is reduced. For further decrease in temperature, $R_{xy}$ quickly reduces to the background value as antiferromagnetic Nd forbids moment formation along the honeycomb element at low temperature. Second, the Hall resistance increases and tends to diverge at higher temperature, $T\geq$ 23 K, in magnetic field application of $H \geq$ 0.25 T. $R_{xy}$ divergence tends to saturate at $H$ $\geq$ 1.75 T. Further increase in field application does not seem to affect Hall resistance anymore. Third, $R_{xy}$ does not exhibit any field dependence above $T\sim$ 32 K, see Fig. 3c. The residual field-independent resistance above $T\sim$ 32 K is most likely arising due to the impurity scattering in the underlying Sn layer. Thus, the phenomenon, above Neel temperature, is limited to a narrow temperature range of 18 K$\leq T \leq$32 K. Interestingly, it is the similar temperature range where Nd-honeycomb was previously demonstrated to manifest the novel Wigner crystal state of magnetic charges.\cite{Yiyao} 

The observed Hall effect around $T\sim$ 20 K does not persist to high temperature or high magnetic field. Therefore, it cannot be a conventional phenomenon. The anomalous observation can be qualitatively explained by the planar Hall effect due to the anisotropic magnetization along honeycomb element, giving rise to the local spin ice order. In principle, the lattice can host equal number of both ‘2-in \& 1-out’ and ‘2-out \& 1-in’ spin ice states. However, such a state of minimum energy is not feasible at finite temperature.\cite{Heyderman} The consequent imbalance in the number of local chirality will generate a net $\Delta \rho$ at a given current, as discussed previously. At much higher temperature, thermal fluctuation is strong enough to destabilize the static moment alignment along the honeycomb element. Accordingly, the intrinsic mechanism behind PHE gradually weakens as temperature increases. We do not observe any planar Hall effect at $T$ = 100 K.

\begin{figure}
\centering
\includegraphics[width=9.5 cm]{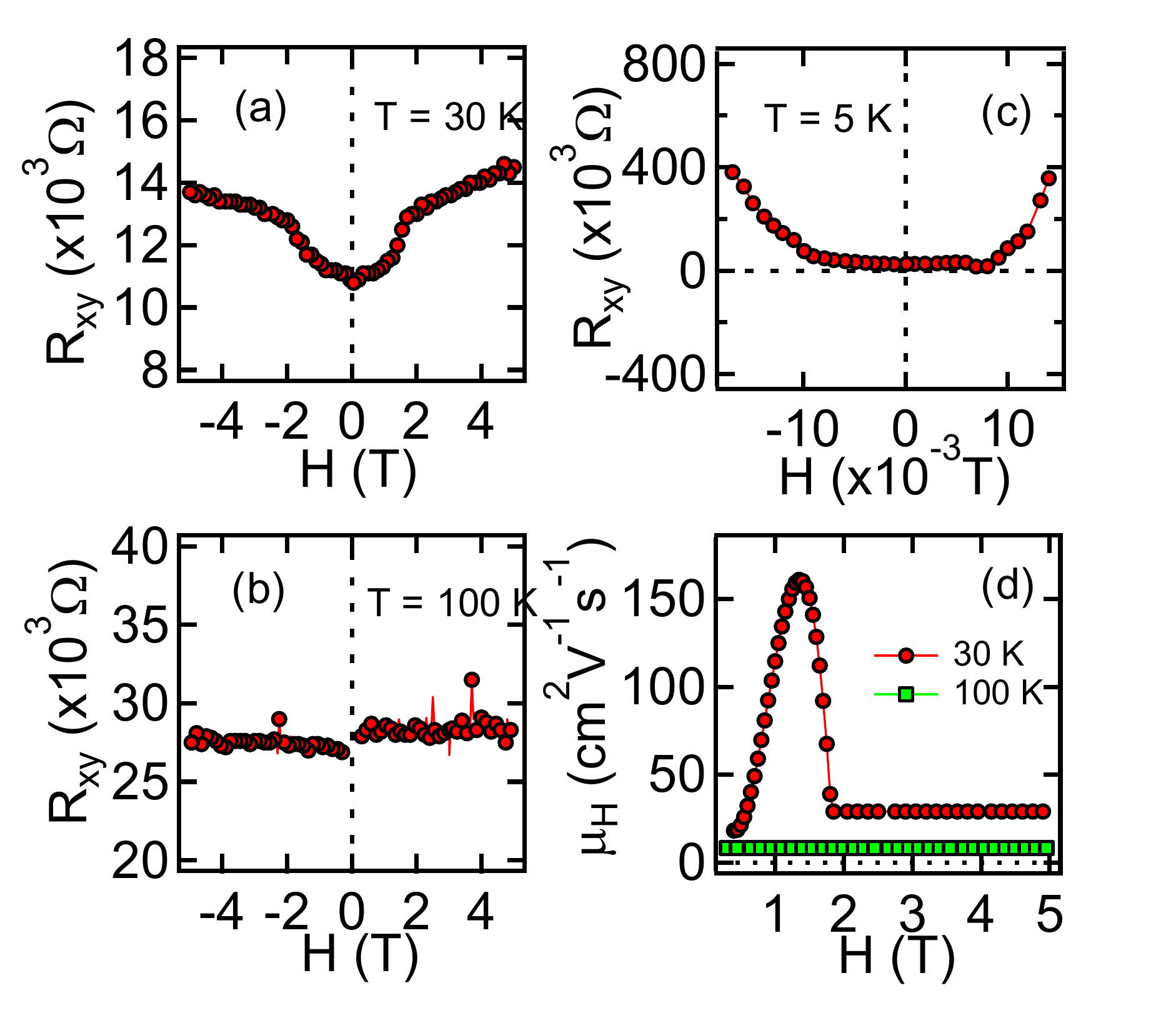} \vspace{-6mm}
\caption{(color online) Field dependence of Hall resistance in Nd-Sn honeycomb. (a--b) Magnetic field dependence of $R_{xy}$ at characteristic temperature of $T$ = 30 K and $T$ = 100 K. As we can see in Fig. (a)  a large enhancement in $R_{xy}$ is detected in modest perpendicular field application at $T$ = 30 K, suggesting planar Hall effect in the system. At higher field, $R_{xy}$ exhibits a very weak linear behavior, which could be arising due to conventional Hall effect. Unlike at $T$ = 30 K, almost no magnetic field dependence in Hall resistance is detected at $T$ = 100 K. Thus, the effect is limited to low temperature and low field. (c) $R_{xy}$ vs. $H$ at $T$ = 5 K. $R_{xy}$ increases by more than two orders of magnitude at $H$ = 0.015 T. (d) Plot of estimated Hall mobility, $\mu_H$, as a function of field at $T$ = 30 K and 100 K. Hall mobility at $T$ = 30 K exhibits strong field dependence. $\mu_H$ saturates above $H\sim$ 1.75 T. $\mu_H$ at $T$ = 100 K does not exhibit any field dependence. Additionally, the Hall mobility at $T$ = 100 K is barely distinguishable from electronic background.
} \vspace{-6mm}
\end{figure}

Magnetic field dependence of the Hall resistivity is somewhat complex, compared to temperature dependence. At modest field application, the local moment arrangement dictates the Hall effect. As shown in Fig. 4a-b, the plot of $R_{xy}$ vs. $H$ at $T$ = 30 K consists of two components: a sharp increment at low field, up to $H \sim$ 1.75 T, and a linear field dependence at high field. While the linear term at higher field could be arising due to the conventional Hall effect, the sharp increase in $R_{xy}$ at low field is attributed to the planar Hall effect due to the local moment configuration in the spin ice state in Nd lattice. A perpendicular field application, larger than $H \sim$ 1.75 T, tends to overcome the shape anisotropy, necessary for the moment formation along the honeycomb element. In addition to the large field application, the effect also disappears at high temperature of $T$ = 100 K where thermal fluctuation is strong enough to destroy the local spin ice moment arrangement. As expected, the experimental data at $T$ = 100 K does not exhibit any field dependence. An entirely different phenomenology develops at low temperature. We observe large Hall resistance at $T$ = 5 K at a very modest magnetic field of $H$ = 0.015 T, see Fig. 4c. In the AFM state at low temperature, Nd lattice does not develop the local spin ice order due to the lack of moment formation. Hence, the planar Hall effect disappears. At the same time, however, a perpendicular field tends to align atomic Nd moment to the field direction, see schematic Fig. 1d. Weakly interacting Nd spins do not require large field application. Furthermore, the absence of thermal fluctuation at low temperature facilitates the induced field effect at much smaller external field. Consequently, the electric charge carriers in Nd-Sn lattice become subjected to a much larger perpendicular magnetic field due to Nd moments. The effect is reflected by strong enhancement in $R_{xy}$ at $H \sim$ 0.015 T, see Fig. 4c. Unlike the diverging nature of Hall resistance at $T \sim$ 25 K at $H$ = 0.5 T or higher, only finite value is detected at low temperature and in low field. It suggests that the PHE generated by the spin ice moments ($\sim$ 10$^{3}$ $\mu_B$) at $T \sim$ 20 K is much stronger than the normal Hall effect due to the algebraically summed individual Nd moments due to $f$-electrons, in the field-induced case at low temperature.

We have also extracted the Hall mobility, $\mu_H$, as a function of magnetic field at different temperatures. The Hall mobility is extracted using the formula of $\mu$(H)=$R$$_{xy}$.$t$/$\rho$.$H$ where $R_{xy}$ is the average Hall resistance at a given field, $\frac{R_{xy}(H)+R_{xy}(-H)}{2}$.\cite{NIST} We show the plot of $\mu_H$ vs $H$ at two characteristic temperatures of $T$ = 30 K and 100 K in Fig. 4d. Two important observations can be inferred from this plot: first, $\mu_H$ ($T$ = 30 K) exhibits a distinct field dependence. We observe that $\mu_H$ first increases significantly as magnetic field increases and attends a maximum value of 150 cm$^2$V$^{-1}$s$^{-1}$ at $H \sim$ 1.5 T. For further increment in field application, $\mu_H$ decreases and saturates at $H \sim$ 1.75 T. We note that the field application above $H \sim$ 1.75 T breaks the shape anisotropy barrier for moment formation, necessary for PHE, along the honeycomb element. Second, $\mu_H$ ($T$ = 100 K) is independent of field. Also, $\mu_H$ ($T$ = 100 K) is very small, barely distinguishable from the electronic background. This could be occurring due to the combination of thermal fluctuation and the enhanced phonon scattering at higher temperature. Clearly, local magnetic moment configuration in the spin ice state plays important role in the Hall mobility, which is significant at $T \sim$ 18--32 K and low field. These observations are consistent with our argument of the planar Hall effect in Nd--Sn honeycomb lattice, which occurs in a narrow temperature and magnetic field range.

In summary, we have performed Hall effect measurements on recently realized artificial honeycomb lattice, made of Nd--Sn connecting elements. The experimental study has revealed planar Hall effect in artificial lattice due to the local spin ice moment configuration in Nd honeycomb. The PHE is observed in a narrow temperature range, around 20 K, and at low magnetic field of $H <$ 1.75 T. Notably, the effect occurs at relatively higher temperature than the anomalous effect typically found in pyrochlore compounds.\cite{Taguchi} At much lower temperature when Nd is in the antiferromagnetic state, modest field application is found to be sufficient to align individual Nd spin to field direction. As a result, strong Hall resistance is detected at low temperature of $T$ = 5 K. Since the latter effect does not involve anisotropic moment formation along honeycomb element, we envisage this to be a typical experimental signature in Hall measurements on any hybrid thin film with clean contact between metallic and magnetic layers. Artificial spin ice of soft magnetic material, such as a permalloy honeycomb lattice, is known to manifest emergent chiral loop ordered state, also called the spin solid state, at low temperature.\cite{Artur,Summers} Future measurements on thermally tunable permalloy honeycomb lattice are highly desirable to further understand the role of artificial moment chirality on planar Hall effect. Artificial magnetic honeycomb lattice created using soft magnetic material can also be utilized to explore the topological Hall effect, as envisaged in nanostructured magnetic lattice,\cite{Vitalii} due to magnetic flux quantization across the chiral loop.

We thank B. Summers and J. Gunasekera for assistance in sample fabrication and electrical measurements. DKS thankfully acknowledges the support by US Department of Energy, Office of Science, Office of Basic Energy Sciences under the grant no. DE-SC0014461. A portion of this research used resources at the Spallation Neutron Source, a DOE Office of Science User Facility operated by the Oak Ridge National Laboratory.
 .

\clearpage


\begin{thebibliography}{99}

\bibitem{Lindberg} O. Lindberg, \textit{Proceeding of The IRE} \textbf{40}, 1414 (1952).

\bibitem{Karplus} R. Karplus, J. M. Luttinger, \textit{Phys. Rev}. \textbf{95}, 1154 (1954).

\bibitem{Harris} S. T. Bramwell, M. J. Harris, B. C. den Hertog, M. J. P. Gingras, J. S. Gardner, D. F. McMorrow, A. R. Wildes, A. L. Cornelius, J. D. M. Champion, R. G. Melko, and T. Fennell, \textit{Phys. Rev. Lett.} \textbf{87}, 047205 (2001).

\bibitem{Pappas} C. Pappas, E. Lelièvre-Berna, P. Falus, P. M. Bentley, E. Moskvin, S. Grigoriev, P. Fouquet, and B. Farago, \textit{Phys. Rev. Lett.} \textbf{102}, 197202 (2009)

\bibitem{Taguchi} Y. Taguchi,1 Y. Oohara,2 H. Yoshizawa,2 N. Nagaosa, Y. Tokura, \textit{Science} \textbf{291}, 2573 (2001)

\bibitem{Neubauer} A. Neubauer, C. Pfleiderer, B. Binz, A. Rosch, R. Ritz, P. G. Niklowitz, and P. Böni, \textit{Phys. Rev. Lett.} \textbf{102}, 186602 (2009)

\bibitem{Ong} M. Hirschberger, R. Chisnell, Y. Lee, and N. P. Ong, \textit{Phys. Rev. Lett.} \textbf{115}, 106603 (2015)

\bibitem{Nagaosa} K. Ohgushi, S. Murakami, N. Nagaosa, \textit{Phys. Rev. B} \textbf{62}, R6065 (2000).



\bibitem{Ye} Jinwu Ye, Yong Baek Kim, A. J. Millis, B. I. Shraiman, P. Majumdar, and Z. Tešanović, \textit{Phys. Rev. Lett.} \textbf{83}, 3737 (1999)

\bibitem{Nagaosa2} N. Nagaosa, J. Sinova, S. Onoda, A. H. MacDonald, and N. P. Ong, \textit{Rev. Mod. Phys.} \textbf{82}, 1539 (2010)

\bibitem{Lee}H. Katsura, N. Nagaosa, and P. Lee, \textit{Phys. Rev. Lett.} \textbf{104}, 066403 (2010)

\bibitem{Potter} T. R. McGuire and R. I. Potter, \textit{IEEE Transactions on Magnetics} \textbf{Mag-11}, 1018 (1975).

\bibitem{Burkov} A. Burkov, \textit{Phys. Rev. B} \textbf{96}, 041110R (2017).

\bibitem{Nitesh} N. Kumar, S. Guin, C. Felser, and C. Shekhar, \textit{Phys. Rev. B} \textbf{98}, 041103R (2018).

\bibitem{Nandy} S. Nandy, G. Sharma, A. Taraphder, and S. Tewari, \textit{Phys. Rev. Lett.} \textbf{119}, 176804 (2017).

\bibitem{Ogrin} F. Ogrin, S. Lee and Y. Ogrin, \textit{J. Mag. Mag. Mat.} \textbf{219}, 331 (2000).

\bibitem{Tang} H. Tang, R. Kawakami, D. Awschalom, and M. Roukes, \textit{Phys. Rev. Lett.} \textbf{90}, 107201 (2003).

\bibitem{Chern} G. W. Chern, \textit{Phys. Rev.} \textbf{Applied 8}, 064006 (2017)

\bibitem{Bak} P. Bak and B. Lebech, \textit{Phys. Rev. Lett.} \textbf{40}, 800 (1978).

\bibitem{Thomale} J. Reuther and R. Thomale, \textit{Phys. Rev. B} \textbf{83}, 024402 (2011).

\bibitem{Taylor} R. C. Taylor, T. R. McGuire, J. M. D. Coey and A. Gangulee, \textit{J. Appl. Phys.} \textbf{49}, 2885 (1978).

\bibitem{Dai} D.-S. Dai, R.-Y. Fang, L.-T. Tong, Z.-X. Lui, Z.-J. Zhou and Z.-H. Lin, \textit{J. Appl. Phys.} \textbf{57}, 3589 (1985).

\bibitem{Heyderman} S. Skjaerrvo, C. Marrows, R. Stamps and L. Heyderman, \textit{Nat. Rev. Phys.} \textbf{2}, 13 (2020).

\bibitem{Shen} Y. Shen, O. Petrova, P. Mellado, S. Daunheimer, J. Cumings and O. Tchernyshyov, \textit{New J. Phys.} \textbf{14}, 035022 (2012).

\bibitem{Mengotti} E. Mengotti, L. J. Heyderman, A. F. Rodriguez, F. Nolting, R. V. Hugli and H.-B. Braun, \textit{Nat. Phys.} \textbf{7}, 68 (2011).

\bibitem{Schiffer} C. Nisoli, R. Moessner and P. Schiffer, \textit{Rev. Mod. Phys.} \textbf{85}, 1473 (2013).

\bibitem{Branford} W. R. Branford, S. Ladak, D. E. Read, K. Zeissler and L. F. Cohen, \textit{Science} \textbf{335}, 1597 (2012).

\bibitem{Russell} S. Park, B. Kim, O. Yavuzcetin, M. Tuominen, T. Russell, \textit{ACS Nano} \textbf{2}, 1363 (2008).

\bibitem{Yiyao} Y. Chen, B. Summers, A. Dahal, V. Lauter, G. Vignale, D. K. Singh, \textit{Adv. Mat.} \textbf{5}, 224413 (2019).

\bibitem{NIST} Information is available at https://www.nist.gov/pml/nanoscale-device-characterization-division/popular-links/hall-effect/hall-effect.

\bibitem{Lauter} V. Lauter, H. Ambaye, R. Goyette, W.-T. Hal Lee and A. Parizzi, \textit{Physica B} \textbf{404}, 2543 (2009).

\bibitem{DWBA} \textit{BornAgain - Software for simulating and fitting X-ray and neutron small-angle scattering at grazing incidence}, http://www.bornagainproject.org, Version 1.15.0

\bibitem{Artur} A. Glavic, B. Summers, A. Dahal, J. Kline, W. Van Herck,  A. Sukhov, A. Ernst and D. K. Singh, \textit{Adv. Sci.} \textbf{5}, 1700856 (2018).

\bibitem{Summers} B. Summers, L. Debeer-Schmitt, A. Dahal, A. Glavic, P. Kampschroeder, J. Gunasekera, D. Singh, \textit{Phys. Rev. B} \textbf{97}, 014401 (2018).

\bibitem{Vitalii} P. Bruno, V. K. Dugaev and M. Taillefumier, \textit{Phys. Rev. Lett.} \textbf{93}, 096806 (2004).

\end{thebibliography}
\end{document}